\documentclass[a4paper]{jpconf}

\usepackage{dcolumn}
\usepackage{bm}

\usepackage[dvips]{graphicx}
\usepackage{wrapfig,subfigure}
\usepackage{amssymb}
\usepackage{color}

 \graphicspath{../Pictures/,../Pictures/Diagram/}

\newenvironment{fig}{\begin{figure}\begin{center}}{\end{center}\end{figure}}
\newenvironment{eqn}{\begin{eqnarray}}{\end{eqnarray}}

\newcommand{\rz}{\rangle}
\newcommand{\lz}{\langle}
\newcommand{\ha}{a}
\newcommand{\had}{a^{\dag}}
\newcommand{\gext}{\kappa}

\newcommand{\DN}{{\delta N_G}}

\newcommand{\Gupm}{\frac{1}{2}\Gamma_{1, \uparrow}^{\rm qp}}
\newcommand{\Gdpm}{\frac{1}{2}\Gamma_{1, \downarrow}^{\rm qp}}
\newcommand{\Goupm}{\frac{1}{2}\Gamma_{\downarrow, 1}^{\rm qp}}
\newcommand{\Godpm}{\frac{1}{2}\Gamma_{\uparrow, 1}^{\rm qp}}

\newcommand{\Gu}{\Gamma_{1, \downarrow}^{\rm qp}}
\newcommand{\Gd}{\Gamma_{1, \uparrow}^{\rm qp}}
\newcommand{\Gou}{\Gamma_{\uparrow, 1}^{\rm qp}}
\newcommand{\God}{\Gamma_{\downarrow, 1}^{\rm qp}}

\newcommand{\diss}{{\rm diss}}
\newcommand{\nbar}{\bar{n}}

\begin{document}

\title{Sub-Poissonian photon statistics in a strongly coupled single-qubit laser}

\author{M. Marthaler, Pei-Qing Jin, Juha Lepp\"akangas and Gerd Sch\"on}

\address{ Institut f\"ur Theoretische Festk\"orperphysik
           and DFG-Center for Functional Nanostructures (CFN), Karlsruher Institut f\"ur Technologie, D-76128 Karlsruhe, Germany}

\ead{mmartha@tfp.uni-karlsruhe.de}

\begin{abstract}
We investigate qubit lasing in the strong coupling limit. The qubit is given by a Cooper-pair box, and
population inversion is established by an additional third state, which can be addressed via quasiparticle tunneling.
 The coupling strength between oscillator and qubit is assumed to be much higher than the quasiparticle tunneling rate. We find that the
photon number distribution is sub-Poissonian in this strong coupling limit. 
\end{abstract}

 The coupling of an oscillator to a superconducting
  qubit \cite{Wallraff2004,wei2006}, to a single-electron transistor \cite{knobel2003,rodrigues2005,armour2004},
   and directly to a tunnel junction
   \cite{mozyrsky2002} are promising possibilities to study quantum effects in
  electromagnetic or mechanical oscillators \cite{huang2003,Connel2010}. On a principle level,
  all of those schemes can be used in multiple ways, for example for
  qubit read-out using the oscillator as detector \cite{Wallraff2005},
 or the qubit can be seen as an artificial atom that can heat and cool the oscillator \cite{Marthaler2008,Andre2010,Marthaler2011}.
 This has been realized in two systems. A flux qubit has been used to to heat and cool an LC-oscillator \cite{hauss2008,Grajcar2008},
  while lasing has been realized using a Cooper-pair box and quasiparticle tunneling to address a third state \cite{Rodrigues2007,Astafiev2007}
 to create population inversion.
  We will investigate the single-qubit laser in the limit of strong coupling \cite{Deppe2010}.
  This regime has an interesting lasing state with sub-Poissonian photon statistics. 
 Our analysis extends previous studies in the strong 
 coupling regime of the micro maser \cite{Pellizzari1994,McKeever} 
 to a system where the atom is permanently coupled strongly to the cavity.

  \begin{figure}
   \begin{center}
 \includegraphics[width=10 cm]{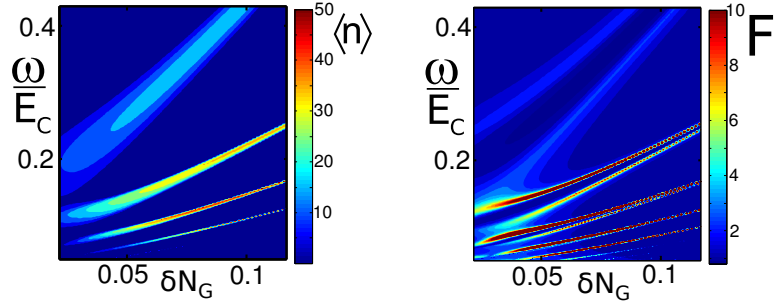}
  \caption{The average photon number $\lz n \rz$,
  the Fano factor $F$ as a function of the frequency
 $\omega$ and the gate charge $\DN$. We see maxima in the
   photon number for the resonance condition $\Delta E=m\omega$,
   where $m=1,2,\ldots$. At resonance, the Fano factor has a minimum
 and the current has a maximum.
   The parameters are $E_J/E_C=0.18$, $\Delta/E_C=2.2$, $eV/E_C=7$,
   $g/E_C=0.01$, $\gext/(E_C/e^2 R_T)=0.1$.}\label{fig:navfanocontour}
   \end{center}
  \end{figure}

  We assume that our qubit is given by a Cooper-pair box coupled via the island charge to the oscillator.
  The qubit eigenstates are a superposition of the
  charge states  $|N=0\rz$ and ${|N=2\rz}$ and are given by 
   \begin{equation}
  |\uparrow \rz =\cos\frac{\phi}{2}|N=2\rz -\sin\frac{\phi}{2}|N=0\rz\,\,\, , \,\,\,
  |\downarrow \rz =\sin\frac{\phi}{2}|N=2\rz+\cos\frac{\phi}{2}|N=0\rz\, .
  \end{equation}
  The rotation angle is given by $\tan \phi=E_J/4 E_C\DN$, where $E_J$ is the Josephson energy
  of the junction, $E_C$ is the charging energy of the island, and $\DN$ is the gate charge. 
  The coherent time evolution of the system is determined by the extended Jaynes-Cummings Hamiltonian ($\hbar=1$)
  \begin{equation}
   H=\frac{1}{2}\Delta E \sigma_z + g(\sigma_z\cos\phi+\sigma_x\sin\phi)(a^{\dag}+a)+\omega a^{\dag}a\, .
  \end{equation}
  A third state $|N=1\rz$  is involved in the lasing cycle,
   but here it is only included on the level of the master equation. 
  In the strong coupling limit the time evolution of the density matrix is
  given in sufficient approximation by a simple balance equation
  \begin{eqn}\label{eq:balance1}
  \dot{\rho}_i=\sum_j\left(\Gamma_{j, i}\rho_j-\Gamma_{i, j}\rho_i\right)\, ,
  \end{eqn}\noindent 
  where $\rho_i=\lz i|\rho|i \rz$, is the probability of the
  system to be in the state $|i\rangle$, 
 $\Gamma_{j, i} \!\!= \!\!\Gamma_{j, i}^{\rm qp} +\Gamma_{j,i}^{\rm diss}$ is the transition rate from  state
  $|j\rz$ to state $|i\rz$ as obtained
  from the Golden rule, 
  \begin{eqnarray}\label{eq:GoldenRulesRatesFull}
  \Gamma_{j, i}^{\rm qp}  =  |\lz
  i|\hat{T}|j\rz|^2I(E_{ji}+e V)\,\,\, ,\,\,\,
  \Gamma_{j, i}^{\rm diss} = \frac{\gext}{\omega}\frac{E_{ji}}{ 1-e^{-E_{ji}/k_B T}}\, 
  |\lz\,i|x|j\rz|^2\,,
  \end{eqnarray}
  where $E_{ji}=E_j-E_{i}$ is the energy difference between initial and final state. We assume linear coupling 
  of the oscillator to reservoir, $x=a^{\dag}+a$, with coupling strength $\kappa/\omega$. 
  The operator $\hat{T}=|1\rz\lz 2|+|0\rz\lz 1|$ decreases the charge of the island by one. 
  The current trough a superconducting junction with resistance $R_T$ is given by
  \begin{eqn}\label{eq:IofV}
  I(E)=\frac{e}{R_T}\int d\omega f(\omega)\left(1-f(\omega+E)\right)N(\omega)N(\omega+E)\,  ,
  \end{eqn}\noindent
  where $N(E)$ is the dimensionless superconducting density of states and
  $f(\omega)$ is the Fermi function at temperature $T$.

  The crucial ingredient for any laser is the creation of population inversion. For the specific example we discuss
  here population inversion is created by quasiparticle tunneling. Without the coupling to the oscillator
  the quasiparticle tunneling rates are given by 
  $\Gamma_{\uparrow, 1}^{\rm qp}\approx\Gamma_{1, \downarrow}^{\rm qp}\propto\sin^2(\phi/2)$, 
  and $ \Gamma_{1 ,\downarrow}^{\rm qp}\approx\Gamma_{\uparrow, 1}^{\rm qp}\propto\cos^2(\phi/2)$ .
  For $\DN>0$ we get $\cos^2(\phi/2)>\sin^2(\phi/2)$. This means
  for the rates that $\Gamma_{1, \uparrow}>\Gamma_{\uparrow,
  1}$ and $\Gamma_{\downarrow, 1}>\Gamma_{1,\downarrow}$. 
  Therefore the  system is most likely to be in the state $|\uparrow\rz$.
  This creates  population inversion in our
  qubit and can be used to generate lasing if the system is coupled to an oscillator.
  The energy for this process is provided by
  the transport voltage $eV$.

  We can numerically calculate the eigenstates as a
  combination of the Fock states of the oscillator $|n\rz$ and the
  eigenstates of the Cooper-pair box $|N=1\rz$,$|\uparrow\rz$, $|\downarrow\rz$. Equation
  (\ref{eq:balance1}) can be solved numerically in the
  stationary case $\dot{\rho}_i=0$. We are interested in the
  average excitation of the oscillator $\lz n\rz=\lz \had\ha\rz$
  and the width of the distribution around this average, which defines
  the Fano factor $F=(\lz n^2\rz-\lz n\rz^2)/\lz n\rz$.
  Results for $T=0$ can be seen in fig. (\ref{fig:navfanocontour}).
  As expected we observe maximal excitation if the
  oscillator is at resonance with the energy difference $\Delta E$.
  The Fano factor at these positions becomes especially
  small, which shows us that the density distribution has a
  sharp peak around the average value of the photon number $\langle n\rangle$.
  Directly at resonance the Fano factor can even be smaller than one,
   which means that we have a sub-Poissonian distribution.
   One can also observe higher order resonances,
   ${\Delta E=m\omega}$, for $m=1,2,\ldots$.  Close to this resonances our 
  master equation is valid for $g\left(g\sqrt{\lz n\rz}/\omega\right)^{m-1}\gg\gext,V/eR_T$,
  where $m$ is the corresponding order of the resonance.
  Only if this condition is fulfilled the
  off-diagonal matrix elements of the density
  matrix can be neglected.

  \begin{fig}
  \includegraphics[width=3.5 in]{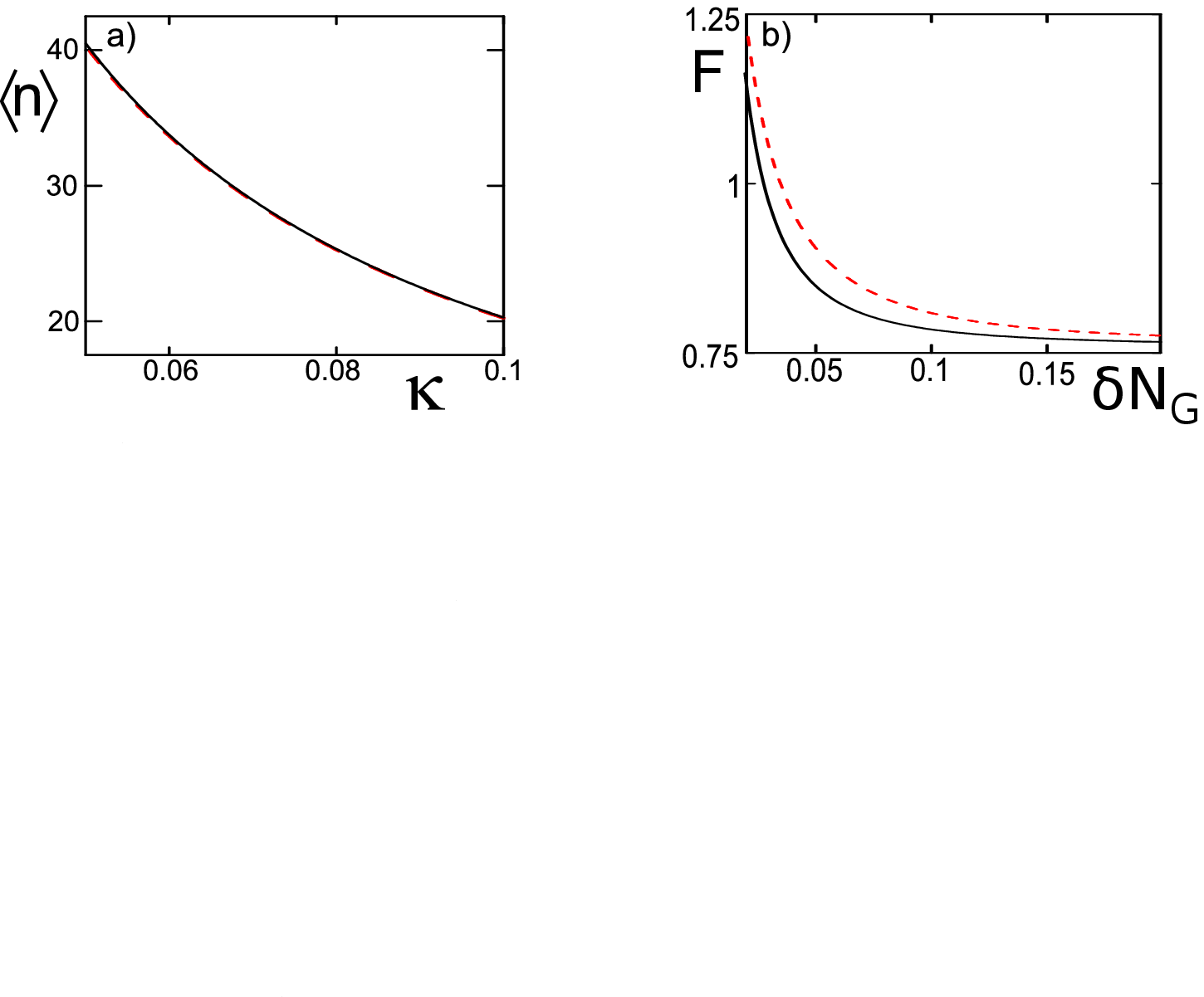}
   \caption{We compare analytical and numerical solutions
 for the average photon number $\lz n \rz$ and the Fano factor $F$. 
 Black line: analytical solution, red-dashed:
   numerical solution.
  a)$\lz n\rz$ as a function of $\gext$. The
    photon number decreases like $\gext^{-1}$. For this plot we have
     chosen $\DN=0.1$. b) The Fano factor F as a function
   of $\DN$. We see that the Fano factor becomes
 smaller than one for large $\DN$. For this plot we have chosen
   $\kappa/(E_C/e^2 R)=0.1$.  For  
  both plots we used the parameters $E_J/E_C=0.18$,
   $\Delta/E_C=2.2$, $eV/E_C=7$, $g/E_C=0.01$.}\label{fig:analytical}
  \end{fig}

  To understand better the emergence of sub-Poissonian
  photon statistics we analyze the system at the
  resonances. On resonance
  the eigenstates are given by the dressed states
  $|\pm,n\rangle= (|\uparrow\rz|n\rz \pm|\downarrow\rz|n+m\rz)/\sqrt{2} $.
  We have now two sets of states, the even states $|\pm,n\rz$ and the odd states
  $|1,n\rz=|N=1\rz|n\rz$. Quasiparticle tunneling leads to
  transitions between even and odd charges, this means
  it will cause transitions between the two sets. Oscillator dissipation
  does not change the charge of a state and therefore only causes
  transitions within each set. We can plug the eigenstates into the
  expression for the transition rates
  (\ref{eq:GoldenRulesRatesFull}). The quasiparticle tunneling rates between the dressed states are given by
  the same rates as tunneling between the eigenstates of the qubit as long as $E_{ij}\ll eV$.
  Assuming that $\lz n \rz$ is large
  we can approximate the oscillator dissipation rates as
  \begin{eqn}\label{eq:analyticalSSETOsciDissipationRates}
  \Gamma_{(1/\pm,n), (1/\pm,n-1)}^{\diss} = (\nbar+1)n\gext \,\, ,\,\,
    \Gamma_{(1/\pm,n), (1/\pm,n+1)}^{\diss} = \nbar\, (n+1)\gext \, ,
  \end{eqn} 
  where $\nbar=(e^{\omega/k_B T}-1)^{-1}$.
  Using theses rates we can
  write the balance equation (\ref{eq:balance1}) as
  \begin{eqn}\label{eq:analyticalbalanceeq1}
   \dot{\rho}_{\pm,n} &=& \Gupm\rho_{1,n}+\Gdpm\rho_{1,n+m}
    +(\nbar+1)(n+1)\gext\rho_{\pm,n+1}
  +\gext\nbar n\rho_{\pm,n-1}\\
 & & -\left(\Goupm+\Godpm+(\nbar+1)n\gext
  +\nbar(n+1)\gext\right)\rho_{\pm,n}\,,\nonumber\\
   \dot{\rho}_{1,n} &=& \sum_{\pm}\left(\Godpm\rho_{\pm,n}+\Goupm\rho_{\pm,n-m}\right)
    -\left(\Gupm+\Gdpm\right)\rho_{1,n}\nonumber\\
  & &  +(\nbar+1)\gext\left((n+1)\rho_{1,n+1}-n\rho_{1,n}\right)
   +\nbar\gext\left(n\rho_{1,n-1}-(n+1)\rho_{1,n}\right)\,.\nonumber
   \end{eqn} 
  We can now find a
  solution for the average oscillator excitation $\lz n \rz$. To
  do this we multiply eqs.
  (\ref{eq:analyticalbalanceeq1}) by $n$ and sum over all $n$.
  We know that for sufficiently low dissipation the distribution
  is peaked around $\lz n\rz\gg 1$, therefore we can neglect $\rho_{i,0}$.
  In the stationary
  limit we get a set of linear equations which
  can be  solved for $n_i=\sum_n n \rho_{i,n}$ ($\{i \in +,-,1\}$).
  The average number of photons is then given in good approximation by 
  $\langle n \rangle=n_+ + n_- + n_1$.
  This yields
  \begin{eqn}\label{eq:analytical_n}
   \lz n\rz &=&\frac{(\God\Gd-\Gou\Gu)m}{\gext\left(\God+\Gou+2(\Gd+\Gu)\right)}+\bar{n}
  \end{eqn} 
  We see that at resonance the oscillator excitations are inversely proportional to the
  dissipation rate $\gext$ and linear in the order of the resonance $m$. The equation we derived here is rather general 
  and describes any strongly coupled three-level laser.
  We can now use the same method to calculate higher moments of
  the photon number $n$. If we multiply equations
  (\ref{eq:analyticalbalanceeq1}) with $n^2$ and sum over all $n$ it
  is straight forward to derive an  approximate equation for $\lz n^2\rz$. 
  For large photon numbers ($\Gou,\Gu \gg \God,\Gd$) the Fano factor reduces to
   \begin{eqn}\label{eq:approx_F}
   F\approx \frac{1}{2}\left(1+ \frac{{\Gou}^2+4 {\Gu}^2 }{(\Gou+2\Gu)^2} \right) +\nbar\,.
   \end{eqn}
   Here we see that at the first order resonance ($m=0$), low temperatures $\nbar\approx 0$
   and for strong coupling, the Fano factor can
   be smaller than one. In fig. \ref{fig:analytical}a) we compare the analytical results
  with the numerics. We have chosen $\DN$ and $\kappa$ such that $\lz n
  \rz$ is large enough to fulfill all approximations we have made
  to derive equations (\ref{eq:analytical_n}) and (\ref{eq:approx_F}). 
  In this case the numerical and analytical results fit
  perfectly.

\end{document}